\documentclass[sigchi]{acmart}
\usepackage{todonotes}
\usepackage[utf8]{inputenc}
\usepackage{url}
\usepackage{hyperref}
\usepackage{color,soul}
\usepackage{array}

\AtBeginDocument{%
  \providecommand\BibTeX{{%
    \normalfont B\kern-0.5em{\scshape i\kern-0.25em b}\kern-0.8em\TeX}}}

\renewcommand\footnotetextcopyrightpermission[1]{} 
\copyrightyear{none}
\acmYear{2019}

\acmConference[RMSE'19]{ACM RMSE Workshop}{September 2019}{Copenhagen, Denmark}

\newcommand{\bumpup}{\vspace{-1.0em}}

\begin{document}

\title{Crank up the volume: preference bias amplification in collaborative recommendation}
\titlenote{Copyright 2019 for this paper by its authors. Use permitted under Creative Commons License Attribution 4.0 International (CC BY 4.0).\\Presented at the RMSE workshop held in conjunction with the 13th ACM Conference on Recommender Systems (RecSys), 2019, in Copenhagen, Denmark.}

\author{Kun Lin}
\authornote{Both authors contributed equally to this research.}
\affiliation{
  \institution{DePaul University}
  \city{Chicago}
  \country{USA}}
\email{linkun.nicole@gmail.com}

\author{Nasim Sonboli}
\authornotemark[1]
\affiliation{%
  \institution{University of Colorado Boulder}
  \city{Boulder}
  \country{USA}}
\email{nasim.sonboli@colorado.edu}

\author{Bamshad Mobasher}
\affiliation{%
  \institution{DePaul University}
  \city{Chicago}
  \country{USA}}
\email{mobasher@cs.depaul.edu}

\author{Robin Burke}
\affiliation{%
  \institution{University of Colorado Boulder}
  \city{Boulder}
  \country{USA}}
\email{robin.burke@colorado.edu}

\renewcommand{\shortauthors}{Kun Lin, et al.}

\begin{abstract}
Recommender systems are personalized: we expect the results given to a particular user to reflect that user's preferences. Some researchers have studied the notion of \textit{calibration}, how well recommendations match users' stated preferences, and \textit{bias disparity} the extent to which mis-calibration affects different user groups. In this paper, we examine bias disparity over a range of different algorithms and for different item categories and demonstrate significant differences between model-based and memory-based algorithms.
\end{abstract}


\maketitle
\section{Introduction}
Recommender systems have become ubiquitous and are increasingly influencing our daily decisions in a variety of online domains. Recently, there has been a shift of focus from achieving the best accuracy \cite{herlocker2004evaluating} in recommendation to other important measures such as diversity, novelty, as well as socially-sensitive concerns such as fairness \cite{hurley2011novelty, jannach2015recommenders}. One of the key issues with which to contend is that biases in the input data (used for training predictive models) are reflected, and in some cases amplified, in the results of recommender system algorithms. This is specially important in contexts where fairness and equity matter or are required by laws and regulations such as in lending (Equal Credit Opportunity Act), education (Civil Rights Act of 1964; Education Amendments of 1972), housing (Fair Housing Act), employment (Civil Rights Act of 1964), with similar provisions in effect in other countries.

The biases in the outputs of recommendation algorithms can be due to a variety of factors in the input data that is fed to the algorithms. As the saying goes: “garbage in, garbage out”. These underlying factors include sample size disparity, having limited features for protected groups, features that are proxies of demographic attributes, human factors or skewed findings \cite{barocas2016big}. These causes are not mutually exclusive and can be present at the same time and they can result in disparate negative outcomes. 

In this paper, we model \textit{bias} as the preferences of users and their tendency to choose one type of item over another. In and of itself, this type of bias is not necessarily a negative phenomenon. In fact, patterns in preference bias are a key ingredient that recommendation algorithms use to construct predictive models and provide users with personalized outputs. However, in certain contexts the propagation of preference biases can be problematic. For example, in the news recommendation domain, preference biases can cause filter bubbles \cite{pariser2011filter} and limit the exposure of users to diversified items. And, in job recommendation and lending domains, existing biases in the input data may reflect historical societal biases against protected groups, which must be accounted for by learning systems \cite{noble2018algorithms}.

Our main goal in this paper is to study how different collaborative filtering algorithms might propagate or amplify existing preference biases in the input data and the different kinds of impact such disparity between input and the output might have on users.
For the purpose of this analysis, we use \textit{bias disparity}, a recently introduced group-based metric\cite{tsintzou2018bias,zhao2017men}. This metric considers biases with respect to the preferences of specific user groups such as men or women towards specific item categories such as different movie genres. This metric evaluates and compares the preference ratio in both the input and the output data and measures the degree to which recommendation algorithms may propagate these biases, in some cases dampening them and in others amplifying them. Throughout this paper we use the notions of preference bias and preference ratio interchangeably.


Our preliminary experiments on a movie rating dataset show that different types of algorithms behave quite differently in the way in which they propagate preference biases in the input data. These findings maybe especially important for system designers in determining the choice of algorithms and parameter settings in critical domains where the output of the system must conform to legal and ethical standards or to prevent discriminatory behavior by the system. As far as we know, this paper is among the first works to have observed this phenomenon in recommendation algorithms. 

We are specifically interested in answering the following research questions: 
\begin{itemize}
    \item \textbf{RQ1} How do different recommendation algorithms propagate existing preference biases in the input data to the generated recommendation lists?
    \item \textbf{RQ2} How does the bias disparity between the input and the output differ for different user groups (e.g., men versus women)?
    \item \textbf{RQ3} How do bias disparity impact individual users with extreme preferences (positive or negative) with respect to particular categories of items?
\end{itemize}

\section{Related work}

As authors in \cite{burke2018balanced} mention, fairness can be a multi-sided notion. Recommender systems often involve multiple stakeholders, including consumers and providers \cite{burke2016towards} and fairness can be sought for for these different stakeholders. In general, fairness is a system goal, as neither side have a good view of the ecosystem and distribution of the resources. Fairness for users/consumers could mean providing similar recommendations to similar users without considering their protected attributes, such as certain demographic features. Methods that seek fairness for consumers of a system fall under the category of consumer-side fairness (C-fairness). Fairness to item-providers (for example sellers on Amazon), may means providing their items a reasonable chance of being exposed/recommended to consumers. This kind of fairness is called the provider-side fairness (P-fairness).

Various metrics have been introduced for detecting model biases. The metrics presented in \cite{yao2017beyond}, such as absolute unfairness, value unfairness, underestimation and overestimation unfairness focus on the discrepancies between the predicted scores and the true scores across protected and unprotected groups and consider the results to be unfair if the model consistently deviates (overestimates or underestimates) from the true ratings for specific groups. These metrics show unfairness towards consumers.

Equality of opportunity discussed in \cite{hardt2016equality} detects whether there are equal proportions of individuals from the qualified fractions of each group (equality in true positive rate). This metric can be used to detect unfairness for both consumers and providers.

Steck \cite{steck2018calibrated} has proposed an approach for calibrating recommender systems to reflect the various interests of users relative to their initial preference proportions. The degree of calibration is quantified using the Kullback-Leibler (KL) divergence. This metric compares the distribution over all the genres of the set of movies played by the user and the same distribution in a user's recommendation list. A post-processing re-ranking algorithm is then used to adjust the calibration degree in the recommendation list.

The authors in \cite{celma2008hits} have discussed another type of bias called popularity bias. Many e-commerce domains exhibit this kind of bias where a small set of popular items, such as those from established sellers, may dominate recommendation lists, while newly-arrived or niche items receive less attention. In this situation, the likelihood of being recommended for popular items will be considerably higher than the rest of the (long-tail) items, potentially resulting in an unfair treatment of some sellers. The methods presented in \cite{abdollahpouri2017controlling, kamishima2014correcting} have tried to break the feedback loop and mitigate this issue. These methods generally try to increase fairness for item providers (P-fairness) in the system by diversifying the recommendation list of users.

The authors in \cite{channamsetty2017recommender} have looked into the influence of algorithms on the output data; they tracked the extent to which the diversity in user profiles change in the output recommendations. \cite{ekstrand2018exploring} has also looked into the author gender distribution in user profiles in the BookCrossing dataset (BX) and has compared it with that of the output recommendations. According to their results, the nearest neighbor methods propagate the biases and strengthen them, and matrix factorization methods strengthen the biases more. Interestingly, our results for matrix factorization methods show the opposite trends possibly indicating the different behavior of algorithms in different domains and datasets.

The work by Tsintzou et al. \cite{tsintzou2018bias} sought to demonstrate unfairness for consumers/users by modeling the bias as the preferences of users. Their proposed metric is called the bias disparity, and is similar in logic to the metric proposed in Steck's work. They both have a user-centric point of view and want to achieve group-fairness. They both calculate the difference between the preference of the user in the input data and the predicted preference of the user by the recommendation algorithm. Bias disparity metric looks at these differences in a more fined-grained way, evaluating the preferences of specific user groups for specific item categories. KL divergence used in Steck's approach measures more generally the difference in preference distributions across genres. The sign value of the bias disparity, on the other hand, gives us information about how input and output biases differ relative to specific categories: negative values indicating the bias has been reversed and positive values indicating it has been amplified. KL divergence, on the other hand, produces non-negative values and cannot differentiate between these two cases.

One of the limitations of the work of Tsintzou et al. \cite{tsintzou2018bias} is that they perform their analysis only for K-nearest-neighbor models. In this paper, we build on their work by considering a variety of recommendation algorithms. We are also interested in understanding how bias affects female and male user groups separately and how it might affect individual users.

\section{Methodology}

\subsection{Bias Disparity}

Let $\mathcal{U}$ be the set of $n$ users and $\mathcal{I}$ be the set of $m$ items and $S$ be the $n \times m$ input matrix, where $S(u,i) = 1$ if user $u$ has selected item $i$, and zero otherwise. 
 
Let $A_{U}$, be an attribute that is associated with users and partitions them into {\it groups} that have that attribute in common, such as gender. Similarly, let $A_{I}$ be the attribute that is associated with items and that partitions the items into {\it categories}, e.g. movie genres.

Given matrix $S$, the input {\it preference ratio} for user group $G$ on item category $C$ is the fraction of liked items by group $G$ in category $C$: 
\begin{equation} \label{eq:1}
    PR_{S}(G, C) = \frac{\sum_{u \in G}\sum_{i \in C} S(u, i)}{\sum_{u \in G}\sum_{i \in \mathcal{I}} S(u, i)} 
\end{equation}
Eq. \eqref{eq:1} is essentially the conditional probability of selecting an item from category $C$ given that this selection is done by a user in group $G$.






The {\it bias disparity} is the relative difference of the preference bias value between the input $S$ and output of a recommendation algorithm $R$, and is defined as follows:

\begin{equation} \label{eq:3}
    BD(G, C) = \frac{PR_{R} (G, C) - PR_{S}(G, C) }{PR_{S}(G, C)}
\end{equation}

We assume that a recommendation algorithm provides each user $u$ with a list of $r$ ranked items $R_{u}$. Let $R$ be the collection of all the recommendations to all the users represented as a binary matrix, where $R(u, i) = 1$ if item $i$ is recommended to user $u$, and zero otherwise. The overall bias disparity for a category $C$ is obtained by averaging bias disparities across all users regardless of the group. For more details on this metric, interested readers can refer to \cite{tsintzou2018bias}.

In this paper, we use bias disparity metric on two levels: 1. {\it Group-based} bias disparity which is calculated based on Eq. \eqref{eq:3} and calculated the bias disparity for two user groups of women and men. 2. {\it General} bias disparity which is also calculated based on Eq. \eqref{eq:3} for all the users in the dataset regardless of their group membership.



Here we assume that $PR_{S} (G, C) > 0$, and $PR_{R} >= 0$. A bias disparity of zero or near zero means that the input and output of the algorithm are almost the same with respect to the prevalence of the chosen category: the algorithm reflects the users' preferences quite closely. A {\it negative} bias disparity means that the output preference bias is less than that of the input. In other words, the preference bias towards a given category is dampened. The extreme value, $BD = -1$, would indicate that a category important in a user's profile is completely missing from the system's recommendations ($PR_{R} = 0$). 
If the bias disparity value is positive, the output preference bias towards an item category is higher than that of the input, indicating that the importance of the given category has been amplified by the algorithm.


\subsection{Algorithms}

The experiments were performed using the \texttt{librec-auto} experimentation platform, \cite{mansoury2018automating}, which is a python wrapper built around the Java-based LibRec \cite{guo2015librec} recommendation library. All experiments were performed using a 5-fold cross validation setting where 80\% of each user's rating data is used for the training dataset and the rest as the test dataset (LibRec's \texttt{userfixed} configuration).

We tested our experiments on four groups of algorithms: memory-based, model-based (ranking), model-based (rating) and baseline. We selected both user-based and item-based k-nearest-neighbor methods from the memory-based category. BPR \cite{rendle2009bpr}, RankALS \cite{takacs2012alternating} were selected from the learning-to-rank category. From the rating-oriented latent factor models \cite{koren2009matrix}, we chose Biased Matrix Factorization (BiasedMF) \cite{paterek2007improving}, SVD++ \cite{koren2008factorization}, and Weighted Regularized Matrix Factorization (WRMF) \cite{hu2008collaborative}. We used a most-popular recommender as a baseline as this algorithm would be expected to maximally amplify the popularity bias in the recommendation outputs.

For each algorithm, we tuned the parameters and picked the one that gives the best performance in terms of normalized Discounted Cumulative Gain (nDCG) of the top 10 listed items. The nDCG values of the algorithms over two experiments in the paper are shown in Table \ref{tab:nDCG}. 

\begin{table}[bth]
\centering
\begin{tabular}{c | c c c}
\hline
\textbf{Algorithm}  & \textbf{Experiment 1} & \textbf{Experiment 2} \\ \hline
MostPopular & 0.480 & 0.460 \\ \hline
ItemKNN & 0.524 & 0.515 \\ \hline
UserKNN & 0.572 &  0.559 \\ \hline
BPR & \hl{\textbf{0.616}} & \hl{\textbf{0.588}} \\ \hline
RankALS & 0.446 & 0.374 \\ \hline
BiasedMF & 0.200 & 0.200 \\ \hline
SVD++ & 0.167 &  0.239\\ \hline
WRMF & 0.507 & 0.498 \\ \hline
\end{tabular}
\caption{nDCG values with selected parameters for the two experiments}
\label{tab:nDCG}
\bumpup\bumpup
\end{table}

\subsection{Dataset}
We ran our experiments on MovieLens 1M\footnote{https://grouplens.org/datasets/movielens} dataset (ML), a publicly available dataset for movie recommendation which is widely used in recommender systems experimentation. ML contains 6,040 users and 3,702 movies and 1M ratings. The sparsity of ratings in this dataset is about 96\%.

\subsection{Experiment Design}

In this section, we look to address these questions: 
\begin{itemize}
    \item What values of the bias disparity are produced by different recommendation algorithms? (RQ1)
    \item Do bias disparity values differ across male and female users in the dataset? (RQ2)
    \item How are users with extreme initial preference ratio effected by bias disparities? (RQ3) 
\end{itemize}

We addressed these questions in three steps: Initially, we selected a subset of the ML dataset consisting of male and female user groups and two movie genres as our item groups. Then, in the {\it first step}, we separately calculated preference ratio (Eq. \ref{eq:1}) of males and females (user groups) on these genres and computed the corresponding the bias disparity values (Eq. \ref{eq:3}). In the {\it second step}, we calculated the preference ratios and bias disparities for our movie genres on the whole user data (without partitioning into separate user groups). In the {\it third step}, we looked into users with zero initial preference ratio on one of the genres to see the effects of different algorithms on bias disparity. Our goal was to determine if input preference ratios were significantly different from the output preference ratios in the recommendations (i.e., if bias disparity was significantly different from 0, due to the dampening or amplification of preference biases). 

In the first step of the experiments, we calculated the group-based bias disparity. As bias disparity represents a form of inaccuracy (users getting results different from their interests), bias disparity differences between groups represent a form of unfairness as the system is working better for some than for others.
 

In the second step of the experiment, we calculated the general bias disparity for the whole population. The comparison of the bias disparity for the whole population (step 2) compared to specific user sub-groups (step 1), can help us understand how algorithms differ in terms of bias disparity across the whole user population.

We ran two sets of experiments, first with Action and Romance genre movies as our item groups, and then with Crime and Sci-Fi genre movies. More details will be mentioned in each experiment.

\section{Experimental Results}

\subsection{Experiment 1: Action and Romance Categories}

In this experiment, we keep the number of items in item groups approximately the same while we create unbalanced user group sizes. The Action and Romance genres are taken as item categories, with 468 and 436 movies in each group respectively. We have 278 women and 981 men for our user groups while each user has at least 90 ratings. After filtering the dataset, we ended up with 207,002 ratings from 1,259 users on 904 items with a sparsity of 18\% for experiment one. 

As we see in Table \ref{tab:ar_input_pref}, the preference ratio of male users is higher for Action genre ($\approx0.70$) compared to the Romance genre ($\approx0.30$) whereas female users have a more balanced preference ratio ($\approx0.50$) over these two movie genres. From comparing the preference ratios of the whole population and sub-groups (table \ref{tab:ar_input_pref}), we observe an overall tendency to prefer the Action genre over Romance genre. This overall bias mainly comes from the preference ratio of the majority male user group.

\begin{table}[!hbtp]
\centering
\begin{tabular}{c c c c}
\hline
\textbf{Genre}  & \textbf{Whole Population} & \textbf{Male}  & \textbf{Female} \\ \hline
Action & 0.675 & 0.721 & 0.502 \\ \hline
Romance & 0.325 & 0.279 & 0.498   \\ 
\hline
\end{tabular}
\caption{Input preference ratio for Action and Romance}
\label{tab:ar_input_pref}
\bumpup\bumpup
\end{table}

\subsubsection{Step 1: Group-based Bias Disparity}
According to the results shown in Figure \ref{fig:ac_pr}, we see that both the neighborhood based methods, UserKNN and ItemKNN, show increased output preference ratio (PR) of both male and female user groups on Action genre by 50\% and around 20\% respectively. While both of these algorithms show increased preference ratio on the Action genre, they have dramatically decreased it for Romance genre, although the preference ratio of women on both genres in the input data were balanced. Accordingly, we see in Figure \ref{fig:ac_bd}, both of these algorithms show negative bias disparities (BD) on Romance for both men and women. 

These results show different outcomes for the two groups because of the different input preference ratio. For the female group, the neighborhood-based algorithms induced a bias towards Action not present in the input; for the male group, the algorithms tend to perpetuate and amplify the existing biases in the input data.


The matrix factorization algorithms show different tendencies. In BiasedMF, the output preference ratio is much lower than the input preference ratio for male users in the Action genre (the opposite of what we observed for the neighborhood-based methods). The PR for the female group is approximately the same. With BiasedMF, the preference ratios of both female and male groups are pushed close to 0.5. We have a negative bias disparity as we see in Figure \ref{fig:ac_bd}, which means that the original preference ratio is underestimated. Interestingly, this algorithm strengthens the bias disparity of both men and women on Romance genre which is an overestimation of their actual preference. We see a similar pattern in SVD++ as well.

WRMF, the other latent factor model, gives inconsistent results from BiasedMF and SVD++. It slightly decreased the preference ratio of women on Action and increased it for men on Action. We see the opposite trend on Romance genre, in other words, the output preference ratio for women on Romance is slightly higher while for men is lower. 

Generally, the absolute value of the BD for the two user groups are not similar. Men have higher absolute values of BD on Romance while women have higher absolute values of BD on Action. As we see in Figure \ref{fig:ac_bd}, different algorithms affect women or men differently. ItemKNN affects the women more than men in both genres, while UserKNN amplifies the bias more for men than women in both genres. BiasedMF and SVD++ increase bias more for men; WRMF, increases the bias slightly more for women.

In this experiment, women had an almost balanced preference over Action and Romance movies, while men prefer Action movies to Romance movies. A well-calibrated algorithm would preserve these tendencies. However, with the influence of the male group, most of the recommender algorithms provide an unbalanced recommendation list specially for women (the minority group). However, BiasedMF and SVD++ run counter to this trend, reversing the bias disparity for both genres. The influence of men's preferences for Action in the overall data is reduced, resulting in fewer unwanted Action movie recommendations for women which is fairer for this group. These two algorithms balance out the exposure of Action and Romance genres for both user groups.



K-nearest-neighbor methods amplify the bias significantly and this behavior could be due to their sensitivity to the popularity bias. Both of the neighborhood-based models show a similar trend to the most-popular recommender (the light blue bar). Romance genre is less favored by the majority group (981 men vs 278 women) in the dataset compared to the Action genre. So, we end up having more neighbors from the majority group as the nearest neighbors (user-knn) or having more ratings from the majority group on a specific genre (item-knn). So, their preference will dominate the preference of the other group on both genres. These methods not only prioritize the preference of the majority group to the minority group, but they also amplify this bias.

\begin{table}[!hbtp]
\centering
\begin{tabular}{|c|c|c|c|}
\hline
\textbf{Algorithm}  & \textbf{Women} & \textbf{Men}  & \textbf{P-value} \\ 
\hline
MostPopular & \textbf{2.519} & 0.659 & 4.29e-05\\ 
ItemKNN     & \textbf{2.100} & 0.994 & 8.89e-25\\ 
UserKNN     & 0.749 & \textbf{1.091} & 2.44e-19\\
BPR         & 0.285 & \textbf{0.678} & 7.79e-03\\
RankALS     & \textbf{0.368}  & 0.306 & \hl{9.99e-01}\\
BiasedMF    & 1.230  & \textbf{2.660} & 1.35e-02 \\ 
SVD++       & 0.803  & \textbf{2.364} & 6.66e-04\\ 
WRMF        & \textbf{0.585}  & 0.523 & 1.76e-05\\
\hline
\end{tabular}
\caption{Bias disparity absolute value sum over Action and Romance}
\label{ar_first_fairness}
\bumpup\bumpup
\end{table}

\subsubsection{Step 2: General Bias Disparity}

\begin{figure}[!hbtp]
    \centering
    \includegraphics[width=\linewidth]{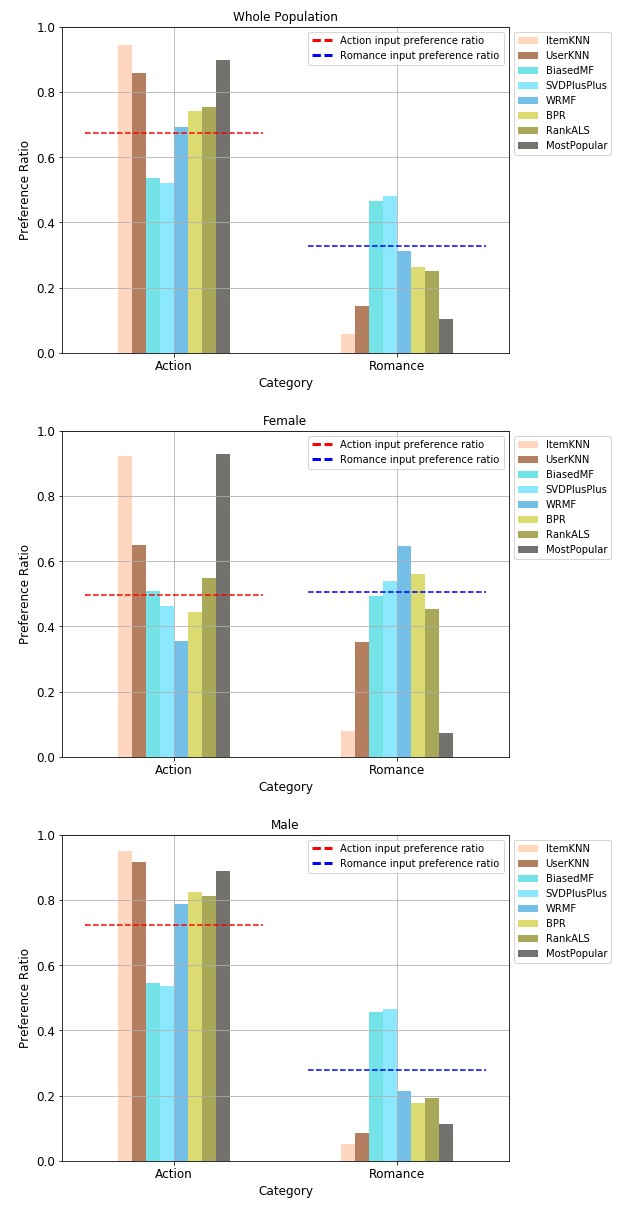}
    \caption{Output Preference Ratio for Action and Romance}
    \Description{Post preference ratio for the action and romance movie, this is for the whole population and the subgroup populations in unbalanced dataset}
    \label{fig:ac_pr}
\end{figure}

In Figure \ref{fig:ac_pr}, the bar shows the preference ratio in the recommendation output and the dashed line shows the input preference ratio for related categories. 

In general, the Action genre (with preference ratio of 0.675) is preferred to the Romance genre (preference ratio of 0.325). As seen in the previous experiment, the two neighborhood-based methods, UserKNN and ItemKNN, both increase the general preference ratio significantly. Our latent factor models (BiasedMF, SVD++, WRMF) show different effects on the preference ratio. None of the matrix factorization algorithms significantly increase the original input preference ratio in the Action genre. BiasedMF and SVD++ significantly decreases the output preference ratio in Action genre, while WRMF keep the the output preference ratio close to the initial preference. 

\begin{figure}[!hbtp]
    \centering
    \includegraphics[width=\linewidth]{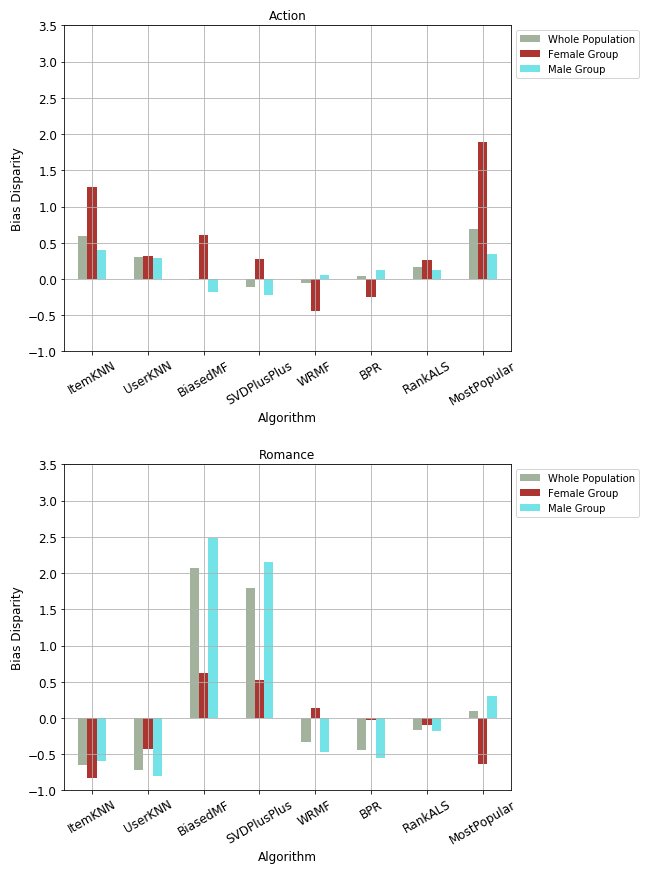}
    \caption{Bias disparity for Action and Romance}
    \Description{Bias disparity for the action and romance movie, this is for the whole population and the subgroup populations in unbalanced dataset}
    \label{fig:ac_bd}
\end{figure}

The Romance category has lower input preference ratio than Action genre, which means that in the input dataset, the population on average prefers Action to Romance. The output preference ratios for this genre show a reverse pattern compared to the Action genre. The neighborhood-based algorithms decrease the preference ratio and most of the matrix factorization algorithms don't change the preference ratio by much except for BiasedMF and SVD++, which significantly increases the preference ratio. We can see the bias disparity change in Figure \ref{fig:ac_bd} as well.


\subsubsection{Step 3: Users with Extreme Preferences}

\begin{figure}[!hbtp]
    \centering
    \includegraphics[width=\linewidth]{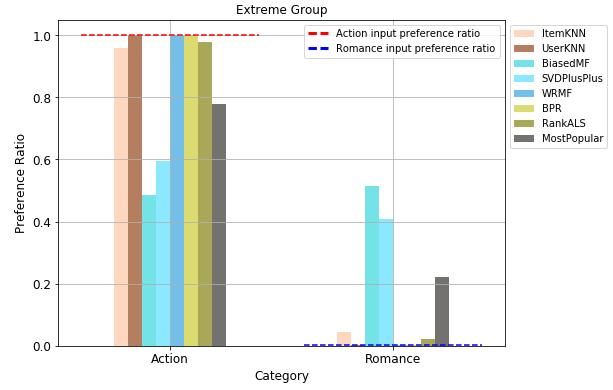}
    \Description{Post preference ratio for the action and romance movie, this is for the whole population and the subgroup populations in unbalanced dataset of the extreme group}
    \caption{Output PR for users with extreme preferences for Action and Romance}
    \label{fig:ac_eg}
\end{figure}

To examine extreme preference cases, we concentrated on users with very low preference ratios across the genres we studied (We excluded the users that had a zero preference ratio on both genres). There were 10 men who had zero preference ratios on the Romance genre, which means that they only watched Action movies. In Figure \ref{fig:ac_eg}, it shows the preference ratio in the recommendation. Some algorithms, like UserKNN, BPR, and WRMF, recommend all Action movies, which is totally consistent with these users' initial preference. Other algorithms, including BiasedMF and SVD++, de-amplify the effects of the preference and show a more diverse recommendation set. When analyzing the preference ratio of the extreme group, the effects of some algorithms become more clearer because of the consistency of the general population and extreme group. 


\subsection{Experiment 2: Crime and Sci-Fi}

In this experiment, our item groups were Crime and Sci-Fi, with 211 and 276 movies in each group respectively. The number of users in both user groups were still unbalanced, 259 female users and 1,335 male users. All of the users had at least 50 ratings from both genres which leaves us with 37,897 ratings from the 1,594 users on 487 items. The sparsity of the dataset was around 95\%.

As it is shown in table \ref{tab:cs_input_pref}, the preference ratios of both male and female users in Crime and Sci-Fi movies are similar. Both men and women have a preference ratio of around 0.7 on Sci-Fi and around 0.3 for the Crime genre. According to Table \ref{tab:cs_input_pref}, the whole population prefers Sci-Fi movies to Crime movies, and we see a similar trend in both user groups, male and female. 

\begin{table}[!hbtp]
\centering
\begin{tabular}{c c c c}
\hline
\textbf{Genre}  & \textbf{Whole Population} & \textbf{Male}  & \textbf{Female} \\ \hline
Crime     &  0.317 & 0.302 & 0.334 \\ \hline
Sci-Fi     & 0.683 & 0.698 & 0.666   \\ \hline
\end{tabular}
\caption{Input Preference Ratio for Crime and Sci-Fi}
\label{tab:cs_input_pref}
\bumpup\bumpup
\end{table}

\subsubsection{Step 1: Group-Based Bias Disparity}

Overall, the group-based bias disparity is very similar to the pattern seen in the whole population. Based on patterns shown in Figure \ref{fig:cs_pr}, the difference between the patterns that we see in Crime genre for both men and women is minimal, and the same trend is true for Romance genre.
The difference in the absolute values of bias disparity between groups is not as enormous as the difference that we saw in Action and Romance (Figure \ref{fig:ac_bd}), which is partly because the two groups have similar preference over the two categories. 

Neighborhood-based algorithms amplify the existing preference bias for both groups. The matrix factorization algorithms either dampen the input bias, like BiasedMF and SVD++, or they don't change the input preference ratio significantly, like WRMF. 

\begin{table}[!hbtp]
\centering
\begin{tabular}{|c|c|c|c|}
\hline
\textbf{Algorithm}  & \textbf{Women} & \textbf{Men}  & \textbf{P-value} \\ \hline
MostPopular & 0.740 & \textbf{0.951} & 0.73\\ 
ItemKNN     & \textbf{1.137} & 0.898 & 0.08\\ 
UserKNN     & \textbf{0.818} & 0.792 & 0.64\\
BPR         & 0.357 & \textbf{0.400} & 0.54\\
RankALS     & \textbf{1.126} & 1.114 & 0.93\\
WRMF        & 0.247  & \textbf{0.311} & 0.15\\
BiasedMF    & 3.089  & \textbf{3.636} & 0.39\\ 
SVD++       & 2.394  & \textbf{2.778} & 0.41\\ 
\hline
\end{tabular}
\caption{Bias disparity absolute value sum over Crime and Sci-Fi}
\label{ar_first_fairness2}
\bumpup\bumpup
\end{table}

\subsubsection{Step 2: General Bias Disparity}

\begin{figure}[h]
    \centering
    \includegraphics[width=\linewidth]{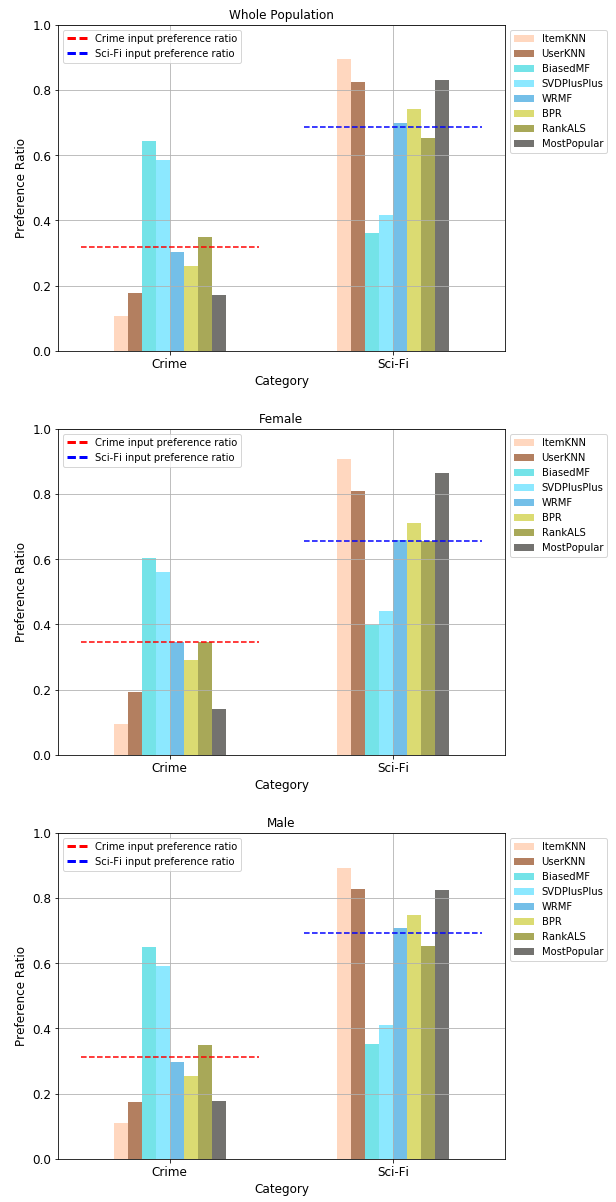}
    \caption{Output preference ratio for Crime and Sci-Fi}
    \Description{Output preference ratio for the Crime and Sci-Fi movie, this is for the whole population and subgroup populations in unbalanced dataset}
    \label{fig:cs_pr}
\end{figure}

As shown in the Figure \ref{fig:cs_pr}, the pattern of Crime and Sci-Fi over the whole population is consistent with Action and Romance. The neighborhood based algorithms, UserKNN and ItemKNN, show an increased output preference ratio for the more preferred genre (Sci-Fi), and a decreased PR for the less preferred genre (Crime). The matrix factorization algorithms show different patterns from neighborhood based algorithms but very similar pattern to experiment 1. BiasedMF and SVD++ have the most significant effects on the preference ratio, increasing the preference ratios of the less favored category and decreasing those of the more favored category. WRMF shows good calibration here. 

\begin{figure}[h]
    \centering
    \includegraphics[width=\linewidth]{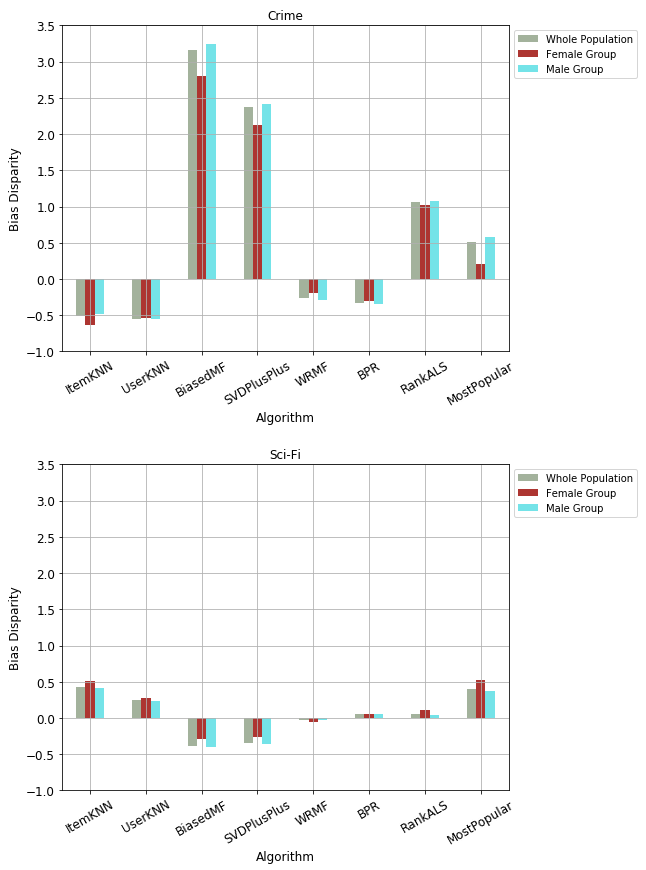}
    \caption{Bias disparity for Crime and Sci-Fi}
    \Description{Bias disparity for the Crime and Sci-Fi movie, this is for the whole population and subgroup populations in unbalanced dataset}
    \label{fig:cs_bd}
\end{figure}

The bias disparity showing in Figure \ref{fig:cs_bd} is also consistent with the bias disparity shown in the Figure \ref{fig:ac_bd} of experiment one.

\subsubsection{Step 3: Users with Extreme Preferences} 
\begin{figure}[!hbtp]
    \centering
    \includegraphics[width=\linewidth]{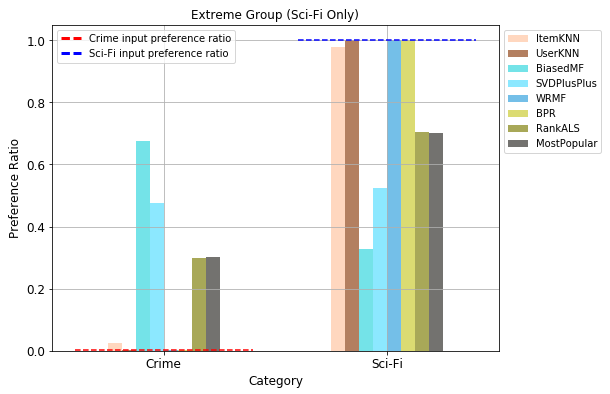}
    \caption{Output Preference Ratio for Crime and Sci-Fi of Extreme Group}
    \Description{Post preference ratio for the crime and sci-fi movie, this is for the whole population and the subgroup populations in unbalanced dataset of the extreme group}
    \label{fig:cs_eg}
\end{figure}

We had 37 users with preference ratio value of zero on Crime movies, meaning that they only watched Sci-Fi movies. The trends that we see in Figure \ref{fig:cs_eg} for this group is pretty similar to Figure \ref{fig:ac_eg}


Similarly to experiment 1, algorithms such as UserKNN, BPR, and WRMF, provide the recommendations well-calibrated to the users' initial preferences whereas BiasedMF and SVD++, significantly ampen the initial preference biases. 


\section{Conclusion and Future Work}

Although we focused here on a handful of the more common movie genres, some important patterns can be seen. Both of the neighborhood-based models show a similar trend towards popularity, consistent with the findings of \cite{jannach2015recommenders}. With these models, we might expect that a dominant group would contribute more neighbors in recommendation generation and would influence predictions by virtue of its presence in these groupings. These methods not only prioritize the preference of the dominant group, but they also amplify the biases for the dominant group across all users.



Different from the previous research on the bias amplification of matrix factorization methods \cite{ekstrand2018exploring}, we observed that different matrix factorization models influence preference biases differently. SVD++ and BiasedMF both dampen the preference bias for different movie genres for both men and women. WRMF algorithm is well-calibrated for the Sci-Fi/Crime genres for both men and women but the behavior is inconsistent for Action/Romance genre.

Each of these model-based algorithms produces a low-rank approximation of the input rating data, but do so in slightly different ways. Jannach et al. \cite{jannach2015recommenders} found that model-based algorithms generally have less popularity bias, so it may be expected that such algorithm would not show as much bias disparity as the memory-based ones. However, further study will be required to understand the interactions between input biases and each algorithm's learning objective. Interestingly, parameter tuning of these algorithms, which produced better accuracy, did not change the bias disparity pattern.


As we have discovered in our experiments, recommendation algorithms generally distort preference biases present in the input data and do so in sometimes unpredictable ways. Different groups of users may be treated in quite different ways as a result. Bias disparity analysis is a useful tool in understanding how aspects of the input data are reflected in an algorithm's output. 



\bibliographystyle{ACM-Reference-Format}
\bibliography{references.bib}
\end{document}